\documentclass{scrartcl}
\usepackage{authblk}
\usepackage{amsmath}
\usepackage{amssymb}
\usepackage{mathrsfs}
\usepackage[T1]{fontenc}
\usepackage{pxfonts}

\usepackage{comment}
\usepackage{color}
\usepackage{bm}
\usepackage{float}
\usepackage{graphicx}
\usepackage{wrapfig}
\usepackage{url}

\title{A dynamic optimal execution strategy under stochastic price recovery}
\date{}
\author[$\dag$]{Masashi Ieda\thanks{ieda@craft.titech.ac.jp}}
\affil[$\dag$]{
	Graduate School of Innovation Management \authorcr
	Tokyo Institute of Technology \authorcr
	2-12-1 Ookayama, Meguro-ku, Tokyo, Japan
}

\begin{document}

\maketitle

\begin{abstract}

In the present paper, we study the optimal execution problem under stochastic price recovery based on limit order book dynamics.
We model price recovery after execution of a large order by accelerating the arrival of the refilling order,
which is defined as a Cox process whose intensity increases by the degree of the market impact.
We include not only the market order but also the limit order in our strategy in a restricted fashion.
We formulate the problem as a combined stochastic control problem over a finite time horizon.
The corresponding Hamilton--Jacobi--Bellman quasi-variational inequality is solved numerically.
The optimal strategy obtained consists of three components:
 (i) the initial large trade;
 (ii) the unscheduled small trades during the period;
 (iii) the terminal large trade.
The size and timing of the trade is governed by the tolerance for market impact depending on the state at each time step,
and hence the strategy behaves dynamically.
We also provide competitive results due to inclusion of the limit order,
even though a limit order is allowed under conservative evaluation of the execution price.

\end{abstract}

\textbf{Keywords:} Optimal execution; Market impact; Limit order book; Price recovery; Hamilton--Jacobi--Bellman quasi-variational inequality

\section{Introduction}
The optimal execution problem, which seeks the optimal way to liquidate or acquire a large asset position,
 is a matter of concern for both market practitioners and academic researchers.
The pioneering work on the optimal execution problem, Bertsimas and Lo \cite{Bertsimas1998},
provides an intuitive optimal strategy to minimize the purchasing cost under a discrete time grid and linear market impact,
specifically dividing the amount of the target asset equally across the time grid.
In seminal papers by Almgren and Chriss \cite{Almgren2001} and Almgren \cite{Almgren2003},
the performance criterion of the execution consists not only of the expected revenue but also a penalty for the uncertainty of the revenue.
In these papers, the market impact is defined by two components: the temporary impact and the permanent impact.
The former work considers a discrete time model and the latter a continuous time model.
The approach using market impact consists of temporary and permanent components in continuous time
 with various risk-return criteria appears in papers such as Schied and Sch\"{o}neborn \cite{Schied2009}, 
 Forsyth \cite{Forsyth2011}, Gatheral and Schied \cite{GATHERAL2011}, Kato \cite{Kato2014}
 and Brigo and Graziano \cite{Brigo2014}.

A recent trend in the study of the optimal execution problem is to include features of the limit order book (LOB).
The shape of the LOB, or how the limit orders stack in the LOB, is a crucial aspect for modeling the market impact.
The resilience of the LOB, which leads the price recovery found after the execution of large orders, is also a significant feature of the LOB.
For studies of the empirical features of the LOB and modelling of the dynamics related to the LOB, we refer the reader to
 Almgren \cite{Almgren2005a}, Large \cite{Large2007}, Bouchaud et. al. \cite{Bouchaud2009},
 Hall and Hautsch \cite{Hall2007}, Muke and Farmer \cite{Mike2008}, Toke \cite{Toke2011}, Gould et. al. \cite{Gould2013}
 and Cont et. al. \cite{Cont2010}.
Research reflecting LOB information in optimal trade execution is addressed in 
Alfonsi and Schied \cite{Alfonsi2010a}, Predoiu et. al. \cite{Predoiu2011}, 
Kharroubi and Pham \cite{Kharroubi2010f}, Guilbaud et. al. \cite{Guilbaud2013b},
and Obizhaeva and Wang \cite{Obizhaeva2013}.
In particular, Obizhaeva and Wang \cite{Obizhaeva2013} study
 linear market impact and exponential-type deterministic price recovery, finding an
optimal strategy consisting of the initial large trade, pre-scheduled small trades during the period, and the terminal large trade.

In the present paper, we study the optimal execution problem under stochastic price recovery based on LOB dynamics.
From the viewpoint of the LOB, the price recovery after execution of a large order is regarded as refilling the stacked limit order at the prior price.
We model price recovery by accelerating the arrival of the refilling order.
Following research on dynamics of the LOB such as \cite{Cont2010} and \cite{Large2007},
 we adopt the Cox process model to describe the arrival of the refilling order.
The intensity of the Cox process is increased by the degree of the market impact.
We define two types of intensity functions, linear and exponential,
and call the former a weak recovery intensity and the latter a strong recovery intensity.

The price recovery possibly induces a waiting time to ease the market impact.
To desterilize this waiting time, we include not only the market order but also the limit order in our execution strategy.
In previous research dealing with the optimal execution problem,
 the limit order is rarely included in the optimal execution strategy as it is an intractable order method compared with the market order.
 Moreover, to the best of our knowledge, the dynamics of the market impact due to a large limit order remains uncertain.
To avoid the complexity due to limit orders, including the market impact, we allow a restriction on the limit order.
Our restricted limit order strategy is regarded as a quasi-iceberg strategy.
We again use a Cox process model to describe the dynamics of the counterpart market order arrivals,
which turns into a Poisson process under our restrictions.

The first main result of the present study is to provide a dynamic execution strategy depending on the state at each time step.
Our new price recovery model, which is defined as a stochastic process, is the source of the dynamic behavior in the strategy.
The key to understanding our strategy is tolerance for the market impact depending on the time remaining and the current state.
The tolerance governs both the size and the timing of the trade.
A quantitative demonstration of the usefulness of including the limit order in the strategy is the second main result.
The proposed optimal strategy gives competitive results in numerical simulations,
 even though the limit order is allowed under conservative evaluation of the execution price.

This paper is organized as follows.
Section 2 introduces the mathematical formulation of our execution problem.
We use the combined stochastic control framework, and
 the market and limit order strategies are formulated by the impulse and regular stochastic controls respectively.
The goal of this section is to derive the corresponding Hamilton--Jacobi--Bellman quasi-variational inequality (HJBQVI).
In Section 3, we present a numerical method for solving the HJBQVI, following a similar procedure to that found in \cite{Ieda2013}.
The HJBQVI is discretized by a finite difference scheme and is converted to an equivalent fixed point problem.
In Section 4, we present numerical results under several conditions, specifically
 the types of recovery intensity functions and either including or excluding the limit order.

\section{Model formulation}

We consider the selling order execution problem: we must sell all share holdings before the terminal time.
Let
$( \Omega, \mathcal{F}, \left\{ \mathcal{F}_t \right\}_{t\geq0}, \mathbb{P})$ be a filtered probability space,
$x_0$ be an initial number of share holdings and let $T$ be the terminal time.
We define our execution problem as the combined stochastic control problem over the finite time horizon $T$. 
We note that we call the number of remaining share holdings as the inventory.

\subsection{Execution Strategy}
We first formulate the market and limit order strategies using impulse and stochastic control strategies respectively.
We note that we cannot use the both limit and market orders at the same time in this framework.

The market order strategy consists of the timing of each execution and the volume of each order.
Let $\{ \tau_j \}_{j>0}$ be a sequence of $\mathcal{F}_t$-stopping times,  that is, the sequence of execution times of market orders,
 and let $\zeta_j$ be the market order volume at time $\tau_j$ defined as an $\mathcal{F}_{\tau_j}$-measurable $\mathcal{Z}_j$-valued random variable,
 where $\mathcal{Z}_j \subset \mathbb{R}^+ \backslash \{0\}$.
We prohibit short selling, that is, $\mathcal{Z}_j$ is bounded by the inventory at time $\tau_j$.
We denote $v=\{\tau_j, \zeta_j\}_{j>0}$, and note that $v$ satisfies the definition of impulse control strategy..

We allow a limit order under the following restrictions:
(i) the order volume is small enough;
(ii) the execution price of the limit order is evaluated as a price that is slightly higher (by the fixed spread size $s$) than the best bid price,
 regardless of the quoted price (see equation (\ref{eq:def_cash}) in Section \ref{sec:inv_cash}). 
Then our limit order strategy consists of a single factor, the limit order volume.
 We denote the control set by $\mathbb{L} \subset \mathbb{R}^+$, which is the set of values we are allowed to choose as the limit order volume.
We describe the arrival of the counterpart order to our limit order by a Poisson process (see, for example, \cite{Cont2010}\footnote{
In the literature, the arrival is modelled by a Cox process whose intensity depends on the spread size.
Our situation coincides with a fixed spread size and thus the intensity is a fixed value.
Hence arrivals are indeed described by a Poisson process.
}).
We denote by $l_t \in \mathbb{L}$, $N^L$ and $\lambda^L$ the limit order volume at time $t$,
 the Poisson process describing the arrival of the counterpart order and the intensity of $N^L$ respectively.

\subsection{Price process and market impact}
We next define processes related to the asset price.
Let $P_t$ be the unaffected part of the asset price process for the market sell order at time $t$,
 and suppose that $P_t$ follows the stochastic differential equation
\begin{align*}
	&\begin{cases}
		dP_t = \sigma(t,P_t) dW_t, \\
		P_0 = p_0 \in \mathbb{R}^+.
	\end{cases}
\end{align*}
We note that $P_t$ is the best bid price, and is not the mid-price.
We denote by $\Xi^{v}_t$ the degree of market impact at time $t$, which is governed by
\begin{align}
	&\begin{cases}
		\displaystyle d\Xi^{v}_t = - \delta_\Xi dN^\Xi_t , \quad \tau_j \leq t < \tau_{j+1},\\
		\Xi^{v}_{\tau_{j+1}} = \Xi^{v}_{\tau_{j+1}^-} + \Gamma(\zeta_{j+1}), \quad j=0,1,2,\cdots, \\
		\Xi^{v}_0 = 0,
	\end{cases}
	\label{eq:def_xi}
\end{align}
where $N^\Xi$ is the Cox process with intensity $\lambda^\Xi(\Xi^{v})$.
The market order increases the market impact through the function $\Gamma$, and the impact recovers following the process $N^\Xi$.

The process $N^\Xi$ is the focus of the present research.
Previous studies describe the dynamics of $\Xi_t$ with a deterministic function.
In contrast, our new model describes the dynamics more precisely in the sense that it includes features of the LOB in the dynamics.
Price recovery is regarded as a random event 
corresponding to the arrival of the limit order in the LOB.
This price recovery model is based on the view that
 (i) a large market order consumes the stacked limit order;
 (ii) the consumed orders at the prior price are refilled by new limit orders;
 (iii) the arrival of the limit order refilling the consumed order is more frequent than that of the usual limit order.
The frequent arrival of the limit order is realized by increasing the intensity $\lambda^\Xi$ in our model.
Empirical studies such as \cite{Large2007} take a similar approach.
 
With the perspective above, we define $\lambda^\Xi(\xi)$ to be an increasing function of $\xi$, and 
 propose the following two candidates: a strong recovery intensity defined by the exponential function
\begin{equation}
\lambda^\Xi(\xi) = \bar{\lambda}_1 \left( e^{\bar{\lambda}_2 \xi} -1  \right),
\label{eq:SRI}
\end{equation}
and a weak recovery intensity defined by the linear function
\begin{equation}
\lambda^\Xi(\xi) = \bar{\lambda}_1  \xi.
\label{eq:WRI}
\end{equation}

\subsection{Inventory and cash holdings}
\label{sec:inv_cash}
Let $X^{w}_t$ be the inventory, the number of remaining shares, at time $t$.
Then $X^{w}_t$ is governed by
\begin{align}
	&\begin{cases}
		\displaystyle dX^{w}_t = - l_t dN^L_t , \quad \tau_j \leq t < \tau_{j+1},\\
		X^{w}_{\tau_{j+1}} = X^{w}_{\tau_{j+1}^-} - \zeta_{j+1}, \quad j=0,1,2,\cdots, \\
		X^{w}_0 = x_0,
	\end{cases}
\end{align}
where $N^L_t$ is a Poisson process describing the arrival of the counterpart order to our limit order.

We denote by $Y^{w}_t$ the cash holdings at time $t$. The dynamics of $Y^{w}_t$ are described by
\begin{align}
	&\begin{cases}
		\displaystyle dY^{w}_t =  l_t(P_t-\Xi_t+s) dN^L_t , \quad \tau_j \leq t < \tau_{j+1},\\
		Y^{w}_{\tau_{j+1}} = Y^{w}_{\tau_{j+1}^-} - \zeta_{j+1}(P_{\tau_{j+1}}-\Xi_{\tau_{j+1}}), \quad j=0,1,2,\cdots, \\
		Y^{w}_0 = 0.
	\end{cases}
	\label{eq:def_cash}
\end{align}
We remark that the profit from sell order execution is evaluated conservatively.
As shown in the second line of equation (\ref{eq:def_cash}),
 the execution price of the market order is assessed as the lowest price affected by the market order.
The execution price of the limit order is appraised slightly higher than the best bid price $P_t+\Xi_t$,
 which is regarded as the minimum price that can be quoted.
Hence the advantage of the limit order under the current setting is primarily avoidance of the market impact.

\subsection{HJBQVI}
We define $J^w_t$, the performance criterion at time $t$, as
\[
J^w_t(x,y,p,\xi) = \mathbb{E} \left[ \left. g(X_T,Y_T,P_T,\Xi_T)  \right| X_t = x, Y_t=y, P_t=p,\Xi_t=\xi  \right],	
\]
where $w=(l,v)$ is the combined stochastic control strategy and $g:\mathbb{R}^+ \times \mathbb{R} \times \mathbb{R} \times \mathbb{R}^+ \rightarrow \mathbb{R}$.
We remark that our performance criterion only depends on the terminal state.
The value function $V_t$ is defined by 
\[
V_t(x,y,p,\xi) := \sup_{w} J^w_t(x,y,p,\xi),
\]
and the corresponding HJBQVI is given by (see e.g. \cite{oksendal2007applied})
\begin{align}
&\max \left( \partial_t V_t(x,y,p,\xi) + \sup_{l \in \mathbb{L}}  \left\{ \mathscr{L}^l V_t(x,y,p,\xi) \right\} ,\right. \nonumber\\
&\hspace*{7em} \left. \sup_{\zeta \in \mathcal{Z}(x) } \left\{ \mathscr{M}^\zeta V_t(x,y,p,\xi) \right\} - V_t(x,y,p,\xi)  \right) = 0
 \label{eq:HJBQVI_org} 
\end{align}
with terminal condition $V_T(x,y,p,\xi) = g(x,y,p,\xi)$,
where $\mathscr{L}^l$ is the infinitesimal generator of the multi-dimensional stochastic process $(X_t,Y_t,P_t,\Xi_t)$ 
\begin{align}
&\mathscr{L}^l V_t(x,y,p,\xi) = \frac{\sigma^2(t,p)}{2} \partial^2_p V_t(x,y,p,\xi)  \nonumber\\
&\hspace*{9em}
	 + \lambda^L \left[ V_t(x-l,y+(p-\xi+s)l,p,\xi) -V_t(x,y,p,\xi) \right]\nonumber\\
&\hspace*{9em}
	 + \lambda^\Xi(\xi) \left[ V_t(x,y,p,\xi-\delta_\Xi) -V_t(x,y,p,\xi) \right],
\end{align}
$\lambda^L$ is the intensity of $N^L_t$, and $\mathscr{M}^\zeta$ is the operator describing the intervention:
\begin{equation}
 \mathscr{M}^\zeta V_t(x,y,p,\xi) :=V_t \left(x- \zeta,y+ \left\{p-\xi-\Gamma(\zeta) \right\} \zeta, p ,\xi-\Gamma(\zeta) \right).
\end{equation}

\section{Numerical method}

\subsection{Terminal wealth criterion and reduced form of the value function}
To solve the HJBQVI numerically, we first specify the form of the function $g$.
Following empirical studies of market impact such as \cite{Almgren2005a},
we use a power law function to describe the market impact:
\begin{equation}
\Gamma(x) = \theta_1 x^{\theta_2},
\end{equation}
where $\theta_1,\theta_2 \in \mathbb{R}^+$.
We define the performance criterion as the terminal wealth including the terminal execution.
The function $g$ is determined by
\begin{equation}
g(x,y,p,\xi) = y+(p-\xi-\Gamma(x))x.
\label{eq:terminal_cond}
\end{equation}
The second term of the above equation represents the terminal execution.

In the current setting, we are able to remove the arguments $y$ and $p$ from the value function
by substituting the ansatz $V_t(x,y,p,\xi) = y+x(p-\xi)+\phi_t(x.\xi)$, so that the HJBQVI (\ref{eq:HJBQVI_org}) becomes 
\begin{align}
&\max \left(
	\partial_t \phi_t(x,\xi) 
   + \sup_{ l \in \mathbb{L} } \left\{ \lambda^L \left[ \phi_t(x-l,\xi) -\phi_t(x,\xi) + ls \right]  \right\}
\right.  \nonumber \\
& \hspace*{7.3em}   + \lambda^\Xi(\xi) \left[ \phi_t(x,\xi-\delta_\Xi) -\phi_t(x,\xi) + x \delta_\Xi \right],
  \label{eq:HJBQVI_red} \\
&\hspace*{5em}\left.
	\sup_{ \zeta \in \mathcal{Z}(x) } \left\{ \phi_t(x-\zeta,\xi+\Gamma(\zeta)) -  x \Gamma(\zeta) - \phi_t(x,\xi) \right\}
\right) = 0 \nonumber
\end{align}
with terminal condition $\phi_T(x,\xi)= -x\Gamma(x)$.

\subsection{Discretization of HJBQVI}
We discretize the HJBQVI (\ref{eq:HJBQVI_red}) using a finite difference scheme with grid size $(\delta_t,\delta_x,\delta_\xi)$:
\begin{align}
&\max \left(
	\frac{\phi_{t_{k+1}} (x_{i_x}, \xi_{i_\xi}) - \phi_{k}(x_{i_x}, \xi_{i_\xi}) }{\delta_t} 
   + \sup_{ l \in \mathbb{L}_\delta } \left\{ \lambda^L \left[ \phi_{t_k} (x_{i_x}-l, \xi_{i_\xi})  -\phi_{t_k}(x_{i_x},\xi_{i_\xi}) + ls \right]  \right\} \right. \nonumber \\
& \hspace*{5em}    + \lambda^\Xi(\xi_{i_\xi}) \left[ \phi_{t_k} (x_{i_x}, \xi_{i_\xi}-\delta_\Xi )  -\phi_{t_k} (x_{i_x}, \xi_{i_\xi}) + x_{i_x}\delta_\Xi \right], 
\label{eq:dis_HJBQVI}\\
&\hspace*{3em}\left.
	\sup_{ \zeta \in \mathcal{Z}_\delta(x_{i_x})} \left\{ \phi_{t_k}(x_{i_x}-\zeta,\xi_{i_\xi}+\Gamma(\zeta)) -  x_{i_x} \Gamma(\zeta) - \phi_t(x_{i_x},\xi_{i_\xi}) \right\}
\right) = 0, \nonumber
\end{align}
where $k,i_x,i_\xi \in \mathbb{N}$, $t_k=k\delta_t$, $x_{i_x}=i_x \delta_x$, $\xi_{i_\xi}=i_{\xi} \delta_\xi$,
and $\mathbb{L}_\delta$ and  $\mathcal{Z}_\delta(x)$ are the discretized sets of $\mathbb{L}$ and $\mathcal{Z}(x)$.
The discretized terminal condition is represented by
\[
\phi_{t_{N^t}} (x_{i_x}, \xi_{i_\xi}) =  -x_{i_x}\Gamma(x_{i_x}),
\]
where $N^t=\frac{T}{\delta_t}$, the other grid numbers are denoted by $N^x=\frac{x_0}{\delta_x}$ and $N^\xi=\frac{g(x_0)}{\delta_\xi}$.
We have assumed that the grid sizes $\delta_t$, $\delta_x$ and $\delta_\xi$ are chosen so that
 the grid numbers $N^t$, $N^x$ and $N^\xi$ are natural numbers.

We define the vector $\left( \phi^k_{(i_x,i_\xi)} \right)_{i_x \in \{0,\cdots,N^x \}, i_\xi \in \{0,\cdots,N^\xi \} } $ by 
$ \phi^k_{(i_x,i_\xi)} =  \phi_{t_k}(x_{i_x},\xi_{i_\xi})$, from which we obtain the HJBQVI  (\ref{eq:dis_HJBQVI}) in matrix form:
\begin{align}
&\max \left(
	\frac{ \phi^{k+1}_{(i_x,i_\xi)} - \phi^k_{(i_x,i_\xi)} }{\delta_t} 
   + \sup_{ l \in \mathbb{L}_\delta} \left\{ \lambda^L \left[ \phi^k_{(i_x-i_l ,i_\xi)} - \phi^k_{(i_x,i_\xi)} + ls \right]  \right\} \right. \nonumber \\
& \hspace*{5em}    + \lambda^\Xi(\xi_{i_\xi}) \left[ \phi^k_{(i_x,i_\xi-1)}  - \phi^k_{(i_x,i_\xi)} + x_{i_x}\delta_\Xi \right], 
\label{eq:mat_QVI}\\
&\hspace*{3em}\left.
	\sup_{ \zeta \in \mathcal{Z}_\delta(x_{i-x})} \left\{ \phi^k_{(i_x-i_\zeta^x, i_\xi+i_\zeta^\xi )} -  x_{i_x} \Gamma(\zeta) - \phi^k_{(i_x,i_\xi)}\right\}
\right) = 0, \nonumber
\end{align}
where 
 $i_l = \lceil \frac{l}{\delta_x} \rceil$,
 $i_\zeta^x = \lceil \frac{\zeta}{\delta_x} \rceil$ and
 $i_\zeta^\xi = \lceil \frac{\Gamma(\zeta)}{\delta_\Xi} \rceil$.

 
We now introduce the constant $h \in \mathbb{R}^+$ satisfying
\[
h > \frac{1}{\delta_t}+2\left( \lambda^\Xi(g(x_0))+\lambda^L \right),
\]
and convert the HJBQVI (\ref{eq:mat_QVI}) into the equivalent fixed point problem
\begin{equation}
\phi^k = \max\left( 
		\sup_{l \in \mathbb{L}_\delta} \left\{ \bar{L}^l \phi^k + \bar{f}^{l,k} \right\},
		\sup_{\zeta \in \mathcal{Z}_\delta(x_{i_x})} \left\{  M^\zeta  \phi^k + K^\zeta \right\}
 \right),
\label{eq:fix_QVI}
\end{equation}
where 
\begin{align*}
	&\bar{L}^l_{ij} = \begin{cases}
		\displaystyle 1 - \frac{1}{h} \left( \frac{1}{\delta_t} + \lambda^\Xi(\xi_{i_\xi}) + \lambda^L \right), \quad i=j=(i_x,i_\xi)\\
		\displaystyle  \frac{\lambda^\Xi( \xi_{i_\xi} ) }{h} ,  \hspace*{3.5cm}  i=(i_x,i_\xi), j=(i_x,i_\xi-1) \\
		\displaystyle  \frac{\lambda^L }{h} , \hspace*{4.3cm} i=(i_x,i_\xi), j=(i_x-i_l, i_\xi )\\
		0, \hspace*{4.5cm} \mathrm{otherwise},
	\end{cases}\\
	&\bar{f}^{l,k}_{(i_x,i_\xi)} =  \frac{1}{h} \left( \frac{1}{\delta_t}\phi^{k+1}_{(ix,i_\xi)} + \lambda^\Xi(\xi_{i_\xi}) x_{i_ix}\delta_\Xi + \lambda^L \right), \\
	&M^\zeta_{ij} = \mathbf{1}_{ \left\{ i=(ix,i_\xi), j=(i_x-i_l, i_\xi+i_\zeta^\xi ) \right\} }, \\
	&K^\zeta_{(i_x,i_\xi)} =  x_{i_x} \Gamma(\zeta).
\end{align*}
We find that, thanks to the factor $h$, $\bar{L}^l$ is a contraction map displayed in matrix form,
 and $M^\zeta$ is a non-expansive map.
We refer to \cite[Section 3.2]{Ieda2013} for the procedure to solve the problem in equation (\ref{eq:fix_QVI}).


\section{Numerical results}
In this section we investigate features of the optimal strategy obtained by the method described in the previous section.
We divide this section into four subsections depending on the type of recovery intensity and the restrictions on limit orders.
Throughout this section, we choose the parameters to obtain the optimal strategy as in Table \ref{tb:param_os}.
\begin{table}[H]
	\centering
	\begin{tabular}{|c|l|c|}
		\hline
		Parameter  & Description           & Value \\ \hline
		$x_0$ & initial inventory & 50    \\
		$\delta_x$ & minimum trading volume & 1    \\
		 & \hspace*{0.5cm} (grid size for $X_t$) &     \\
		$\delta_t$ & size of time step & 0.001 \\
		$\delta_\Xi$ & grid size for $\Xi_t$& 1 \\
		$s$ & spread size  & 1    \\
		$\theta_1$ & amplitude of market impact  & 2    \\
		$\theta_2$ & exponent of market impact  & 1    \\
		$\bar{\lambda}_1$ & amplitude of price recovery  & 1    \\
		$\bar{\lambda}_2$ & exponent of price recovery  & 1    \\
		\hline
	\end{tabular}
	\caption{Parameters related to execution strategy}
	\label{tb:param_os}
\end{table}

The key to understanding the optimal behavior is that
	the tolerance for the market impact depends on the time remaining and inventory.
We are able to capture the intuitive ideas that
	(i) the incentive for quick liquidation using the market order becomes higher over the course of time, especially near the terminal time;
	(ii) a lower inventory and a larger market impact induce a wait for price to recover; and
	(iii) if the limit order is available, it is better to quote the limit order as compared with waiting for the price to recover.

\subsection{Strong recovery intensity without limit order}
\label{sec:num_res_strong}
We first consider the case with strong price recovery intensity as defined in equation (\ref{eq:SRI}),
 and set $T=10$, $\lambda^L=0$ and $l_{\max} = 0$.
Note that we are not allowed to use the limit order strategy under the current setting.

\begin{figure}[H]
	\centering
	\includegraphics[width=15cm]{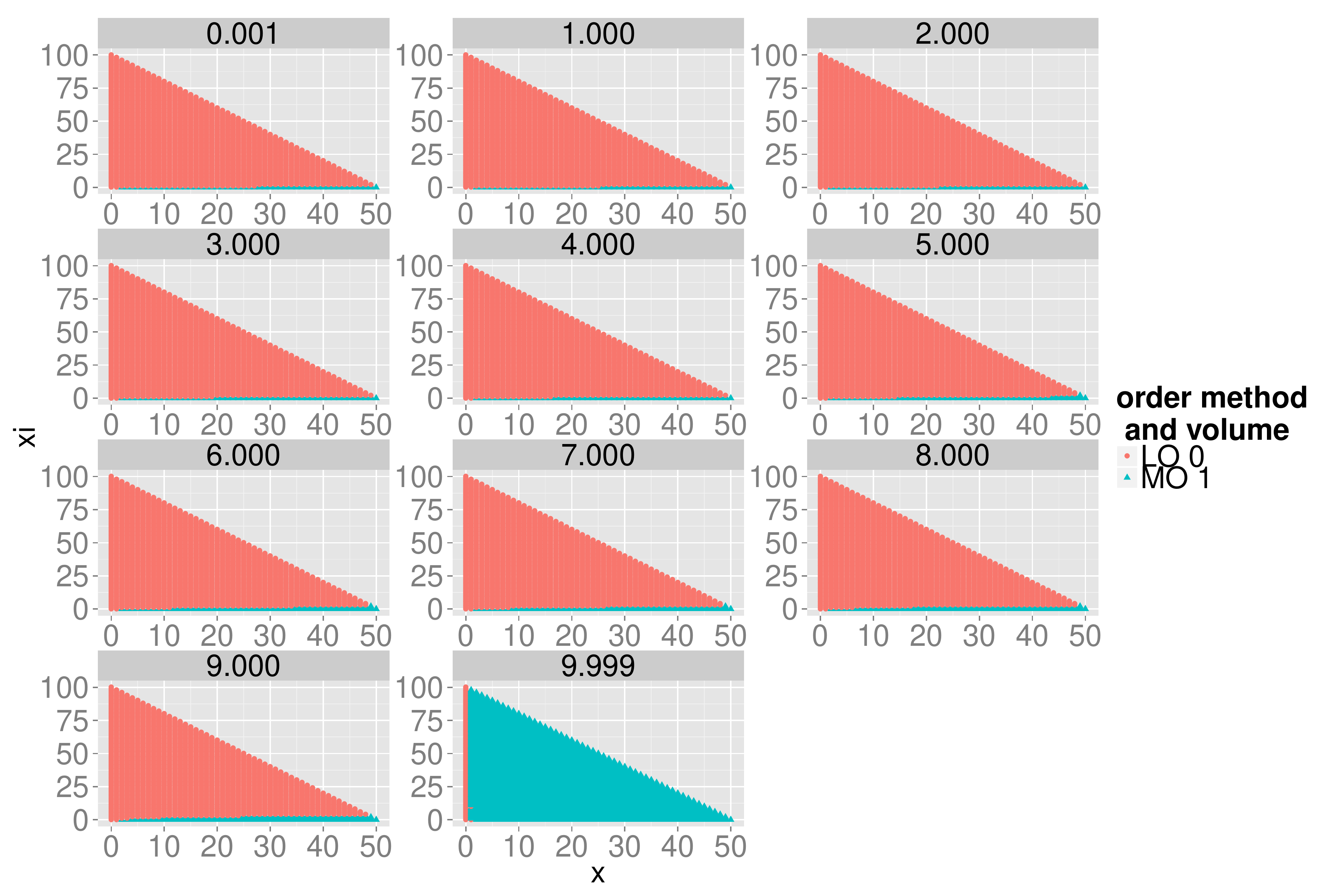}
	\caption{
		The optimal strategy under strong recover intensity without limit order when $T=10$.
		Each panel displays a snapshot at the time indicated above the panel.
		A red dot indicates that waiting for the price to recover is optimal, while
		a blue triangle indicates that selling one share, the minimum trading volume, is optimal.
	}
	\label{fig:strong_t10_stra_ent}
\end{figure}

Figure \ref{fig:strong_t10_stra_ent} shows the optimal strategy, where
each panel displays a snapshot at the time indicated on the top of the panel.
A red dot indicates that waiting for the price to recover is optimal, while
a blue triangle indicates that selling one share, the minimum trading volume, by a market order is optimal.
Because we prohibit short selling, the maximum value of $\Xi_t$ is $g(x_0-X_t)$.
Hence, the optimal strategy is calculated in the triangular regions shown in Figure \ref{fig:strong_t10_stra_ent}.
We observe that the region filled by red dots is dominant except in the last panel, which describes snapshot point close to the terminal time.
We also note that the blue triangles appear in the region where $\Xi_t$ takes a small value.
These results imply that if we observe the large market impact and enough time remains,
then it is optimal to take no action because the price is expected to recover.

\begin{figure}[H]
	\centering
	\includegraphics[width=15cm]{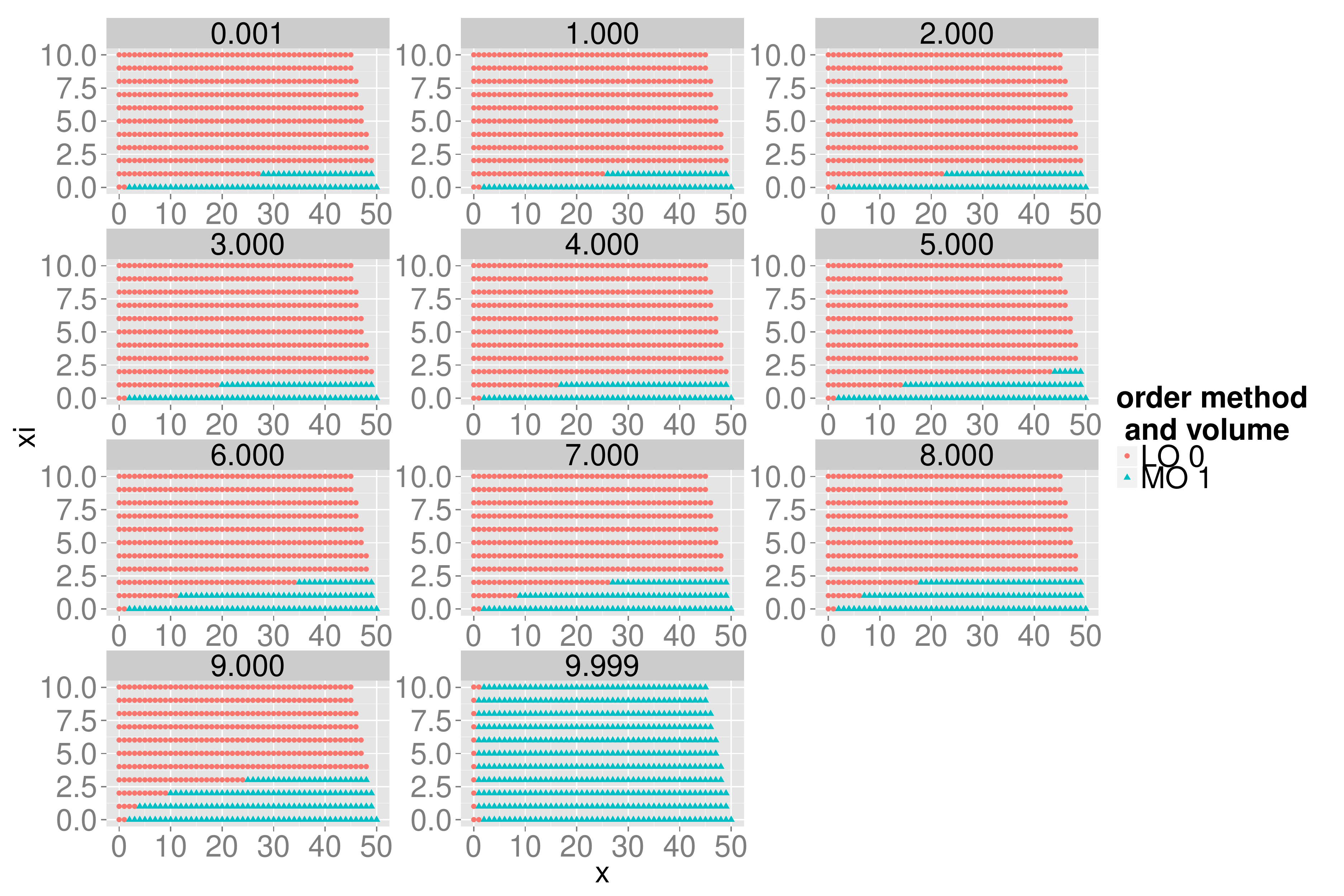}
	\caption{
	The optimal strategy, focusing on the region $\Xi_t \leq 10$
	 under strong recovery intensity without limit order when $T=10$.
	A red dot indicates that waiting for the price to recover is optimal, while a blue triangle indicates that selling one share, the minimum trading volume, is optimal.
	}
	\label{fig:strong_t10_stra}
\end{figure}

We next find that the region filled by blue triangles expands over the course of time.
However, Figure \ref{fig:strong_t10_stra_ent} does not have enough resolution
 to discuss the time evolution of the optimal strategy.
Therefore, we focus on the region where $\Xi_t$ takes a lower value as shown in Figure \ref{fig:strong_t10_stra}.
We find that the region filled by blue triangles expands from the bottom right corner to the top left corner over the course of time.
We also observe that selling one share by a market order is optimal when $\Xi_t=0$, that is, the market impact fully recovers.

The key to understanding the dynamics described in Figure \ref{fig:strong_t10_stra} is tolerance for the market impact.
If the tolerance is high, the optimal strategy is to wait the recovery of the market impact,
 whereas if the tolerance is low, it is optimal to liquidate by market orders.
The region filled by red dots (respectively blue triangles) in Figure \ref{fig:strong_t10_stra} is regarded as the high (resp. low) tolerance region.
The manner of expanding the low tolerance region argues that tolerance for the market impact decreases with an increase in the inventory and over the course of time.
This view is quite consistent with our intuition.

\begin{figure}[H]
	\begin{minipage}[t]{0.48\columnwidth}
		\centering
			\includegraphics[width=7.5cm]{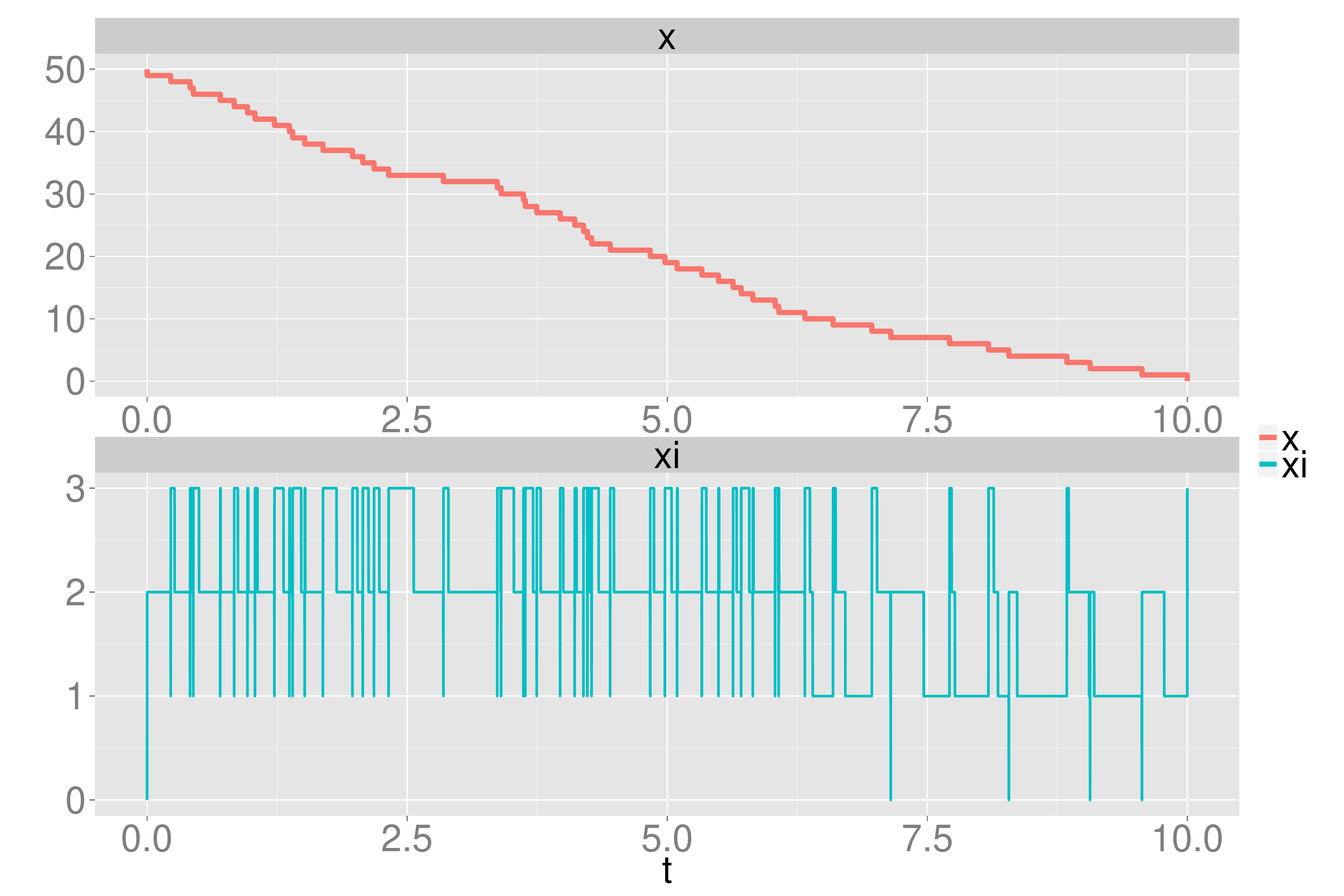}
			\caption{
			A sample path under strong recover intensity without limit orders when $T=10$. 
			This figure describes the fast price recovery case.
			The upper and lower panels display the time evolution of the inventory $X_t$ and the degree of the market impact $\Xi_t$ respectively. 
			}
			\label{fig:strong_sample_T10_fr}
	\end{minipage}
	\hfill
	\begin{minipage}[t]{0.48\columnwidth}
		\centering
			\includegraphics[width=7.5cm]{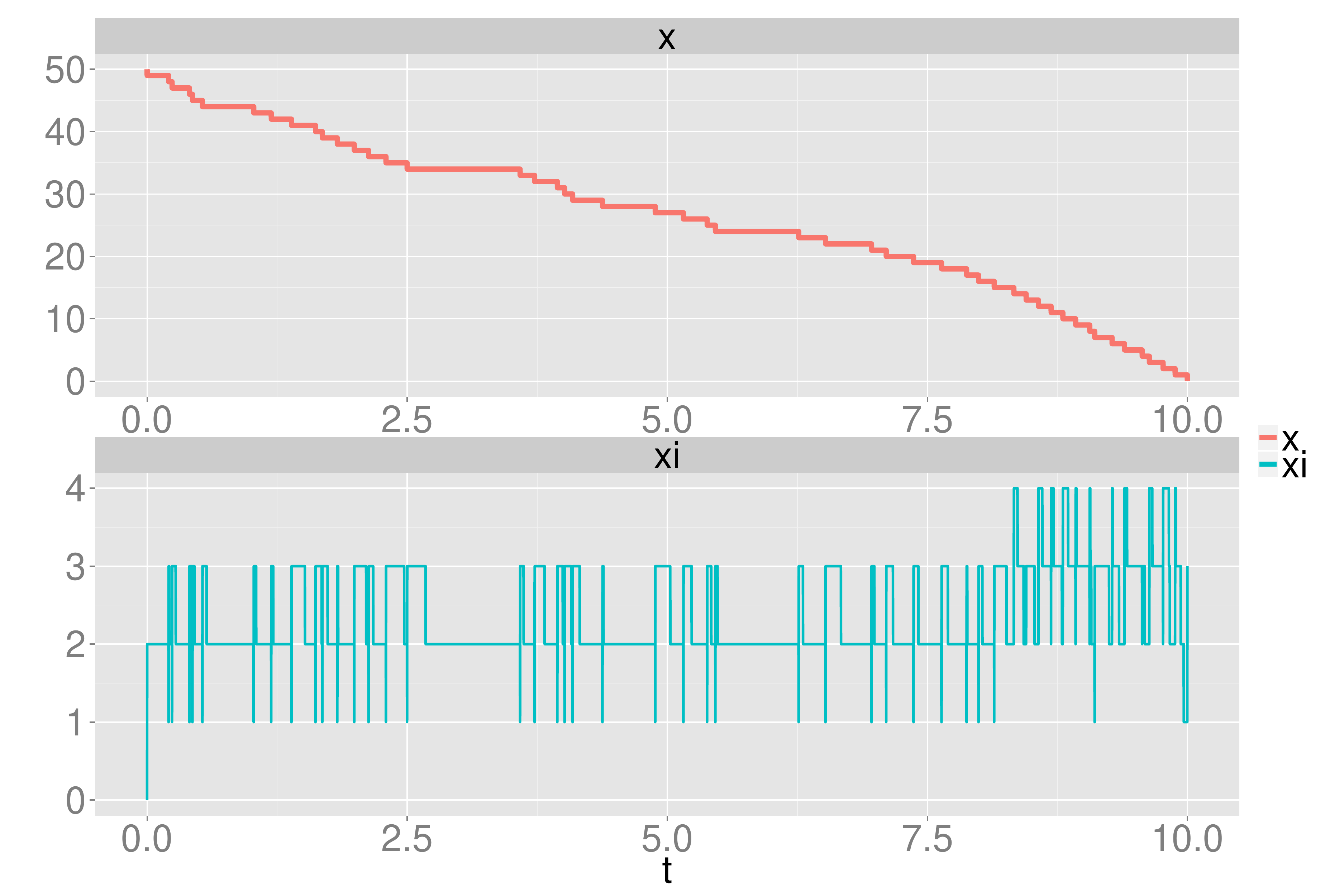}
			\caption{
			A sample path under strong recover intensity without limit orders when $T=10$. 
			This figure describes the slow price recovery case.
			The upper and lower panels display the time evolution of the inventory $X_t$ and the degree of the market impact $\Xi_t$ respectively.
			}
			\label{fig:strong_sample_T10_sr}
	\end{minipage}
\end{figure}

The interpretation using tolerance for the market impact is supported
 by Figure \ref{fig:strong_sample_T10_fr} and Figure \ref{fig:strong_sample_T10_sr}.
These figures display sample paths followed by the numerical simulation:
 the upper and lower panels describe the time evolution of the inventory $X_t$ and the degree of market impact $\Xi_t$ respectively.
We observe that sell order execution by a market order with minimum trading volume occurs when $\Xi_t$ takes a certain value.
The value that triggers execution is $\Xi_t =1$ before $t=7$, after which different behaviors can be detected.
From Figure \ref{fig:strong_sample_T10_fr} and Figure \ref{fig:strong_sample_T10_sr}, we find that the triggers are $\Xi_t=0$ and $\Xi_t=2$ respectively
We are able to explain this phenomenon by considering the remaining inventory.
We observe that $X_t$ in Figure \ref{fig:strong_sample_T10_fr} takes a lower value than in Figure \ref{fig:strong_sample_T10_sr}
 in the region after $t=7$.
Thus the tolerance increases in the sample path shown in Figure \ref{fig:strong_sample_T10_fr} 
 and decreases in the path shown in Figure \ref{fig:strong_sample_T10_sr}.
Focusing on the lower panel in both figures, we find that $\Xi_t$ decreases more frequently in Figure \ref{fig:strong_sample_T10_fr} than it does in \ref{fig:strong_sample_T10_sr}.
The natural behavior is to speed up execution when price recovery is fast, and so there are frequent arrivals of refill orders.

\begin{figure}[H]
	\centering
	\includegraphics[width=15cm]{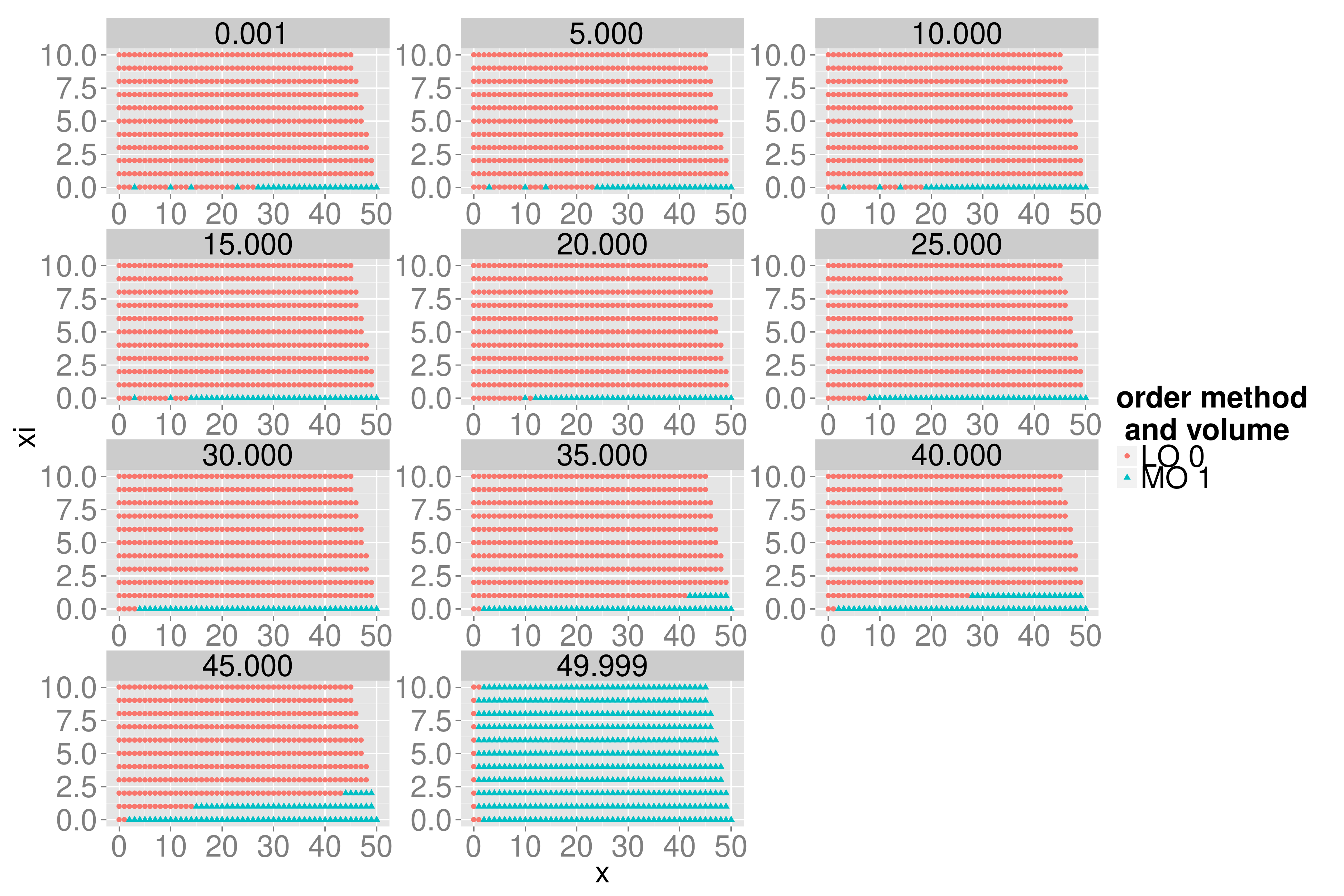}
	\caption{
	The optimal strategy focusing on the region $\Xi_t \leq 10$
	 under strong recovery intensity without limit orders when $T=50$.
	A red dot indicates that waiting for the price to recover is optimal, while a blue triangle indicates that selling one share, the minimum trading volume, is optimal.
	}
	\label{fig:strong_t50_stra}
\end{figure}

\begin{figure}[H]
	\begin{minipage}[t]{0.48\columnwidth}
		\centering
			\includegraphics[width=7.5cm]{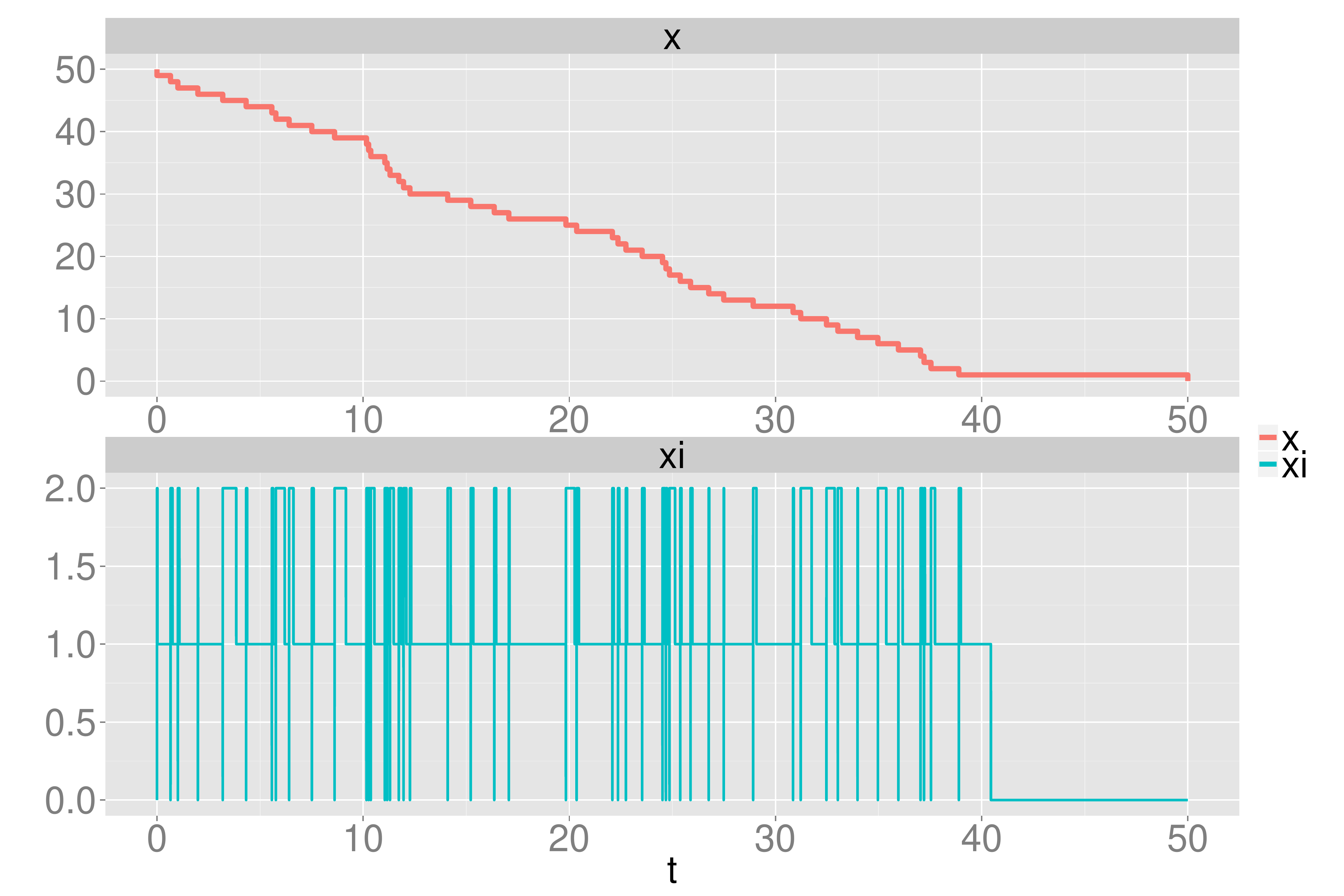}
			\caption{
			A sample path under strong recovery intensity without limit orders when $T=50$. 
			This figure describes the fast price recovery case.
			The upper and lower panels display the time evolution of the inventory $X_t$ and the degree of the market impact $\Xi_t$ respectively.
			}
			\label{fig:strong_sample_T50_fr}
	\end{minipage}
	\hfill
	\begin{minipage}[t]{0.48\columnwidth}
		\centering
			\includegraphics[width=7.5cm]{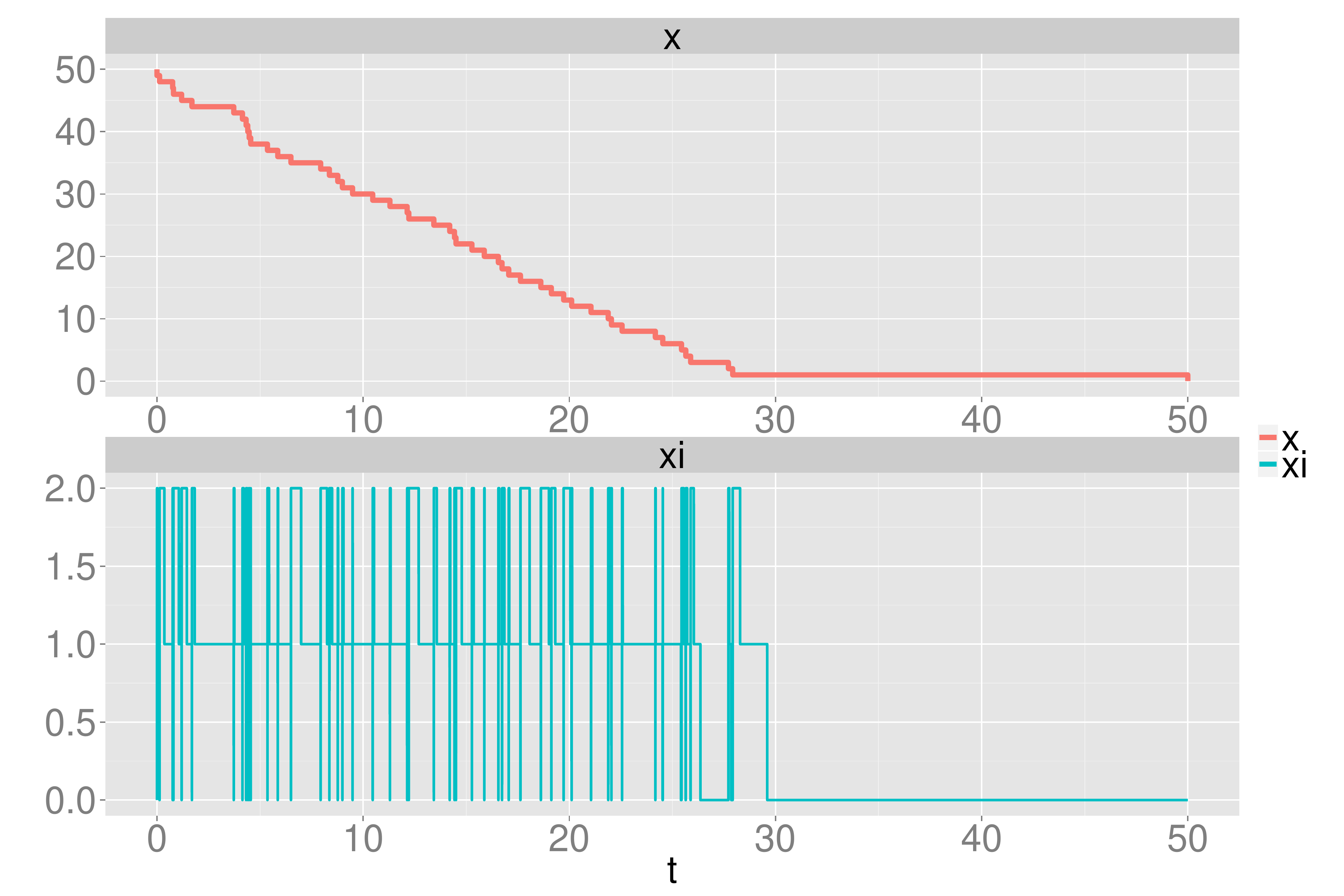}
			\caption{
			A sample path under strong recovery intensity without limit orders when $T=50$. 
			This figure describes the slow price recovery case.
			The upper and lower panels display the time evolution of the inventory $X_t$ and the degree of the market impact $\Xi_t$, respectively.
			}
			\label{fig:strong_sample_T50_sr}
	\end{minipage}
\end{figure}

From the point of view of tolerance for market impact, and the results when $T=10$,
 we predict that
 (i) if we choose a larger value for $T$ then the tolerance will increase; and
 (ii) as a result, the behavior tends to wait for full price recovery over the entire sample paths.
A longer simulation gives results that support this.
Figure \ref{fig:strong_t50_stra} displays the optimal strategy when $T=50$ in the same manner as in Figure \ref{fig:strong_t10_stra}.
The high tolerance region is more dominant compared with that when $T=10$.
The sample paths vividly describe that the optimal strategy is to wait for full price recovery.
We choose a sample path in the fast price recovery case, where price recovery is found frequently, and display it in Figure \ref{fig:strong_sample_T50_fr}.
A sample case of slow price recovery is displayed in Figure \ref{fig:strong_sample_T50_sr}.
Both imply that the optimal behavior coincides with our prediction: waiting for full price recovery is observed over the entire sample path.

\subsection{Weak recover intensity without limit orders}
In this subsection, we show the results using the weak recovery intensity defined in equation (\ref{eq:WRI}),
again excluding limit orders.

\begin{figure}[H]
	\centering
	\includegraphics[width=15cm]{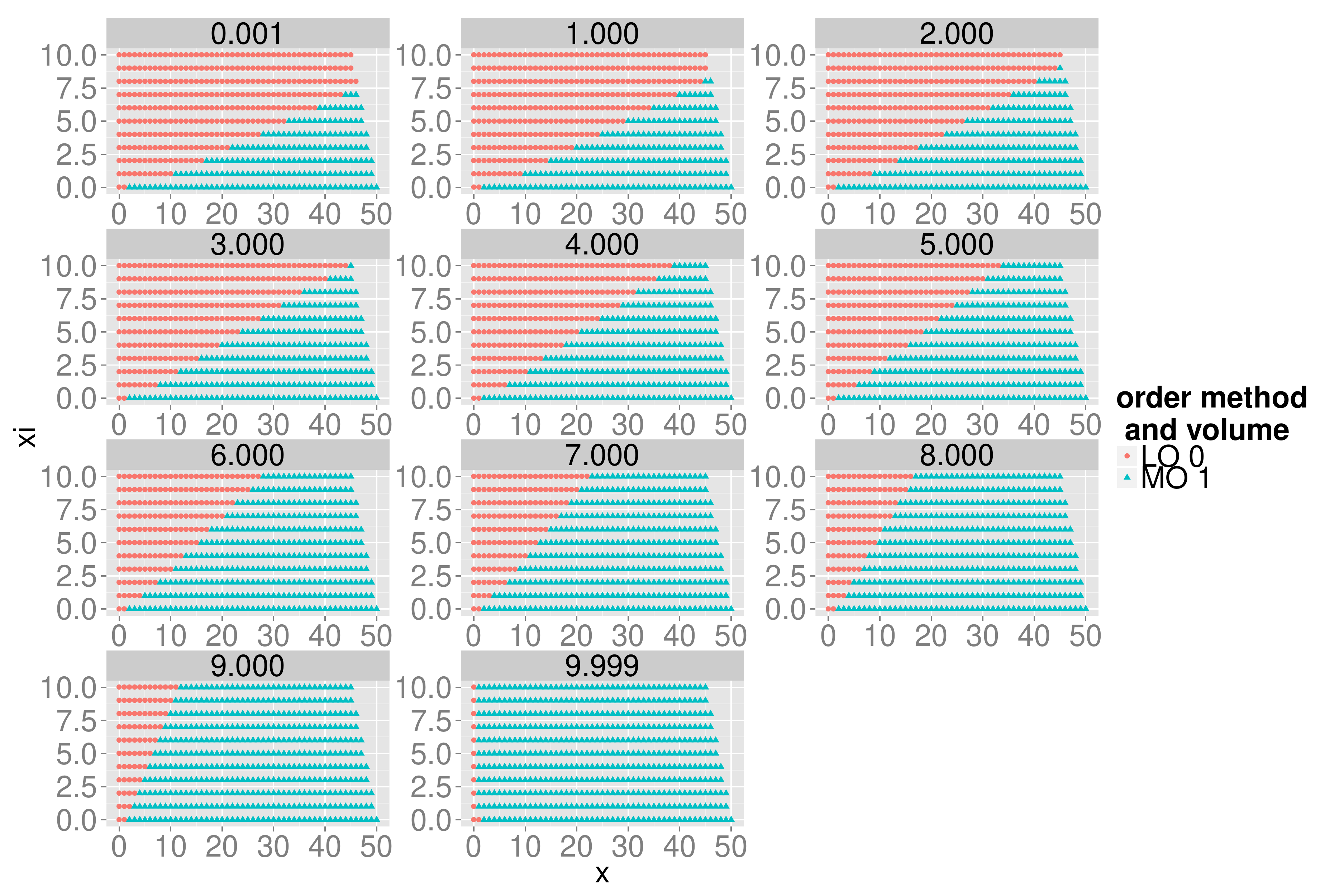}
		\caption{
	The optimal strategy focusing on the region $\Xi_t \leq 10$
	 under weak recovery intensity without limit orders when $T=10$.
	A red dot indicates that waiting for the price to recover is optimal, while a blue triangle indicates that selling one share, the minimum trading volume, is optimal.	}
	\label{fig:weak_t10_stra}
\end{figure}

Let us discuss the results for $T=10$.
Figure \ref{fig:weak_t10_stra} describes the optimal strategy in the same manner as in Figure \ref{fig:strong_t10_stra_ent}.
It is obvious that the triangles representing point with low tolerance for market impact
 occupy a larger region than they do in Figure \ref{fig:strong_t10_stra}.
This result indicates that tolerance for market impact is less than in the case studied in the previous subsection.
The reason is clear: 
because the intensity in equation (\ref{eq:WRI}) implies that refill orders arrive less frequently compared with the intensity in equation (\ref{eq:SRI}),
 we do not expect the price to recover.

\begin{figure}[H]
	\begin{minipage}[t]{0.48\columnwidth}
		\centering
			\includegraphics[width=7.5cm]{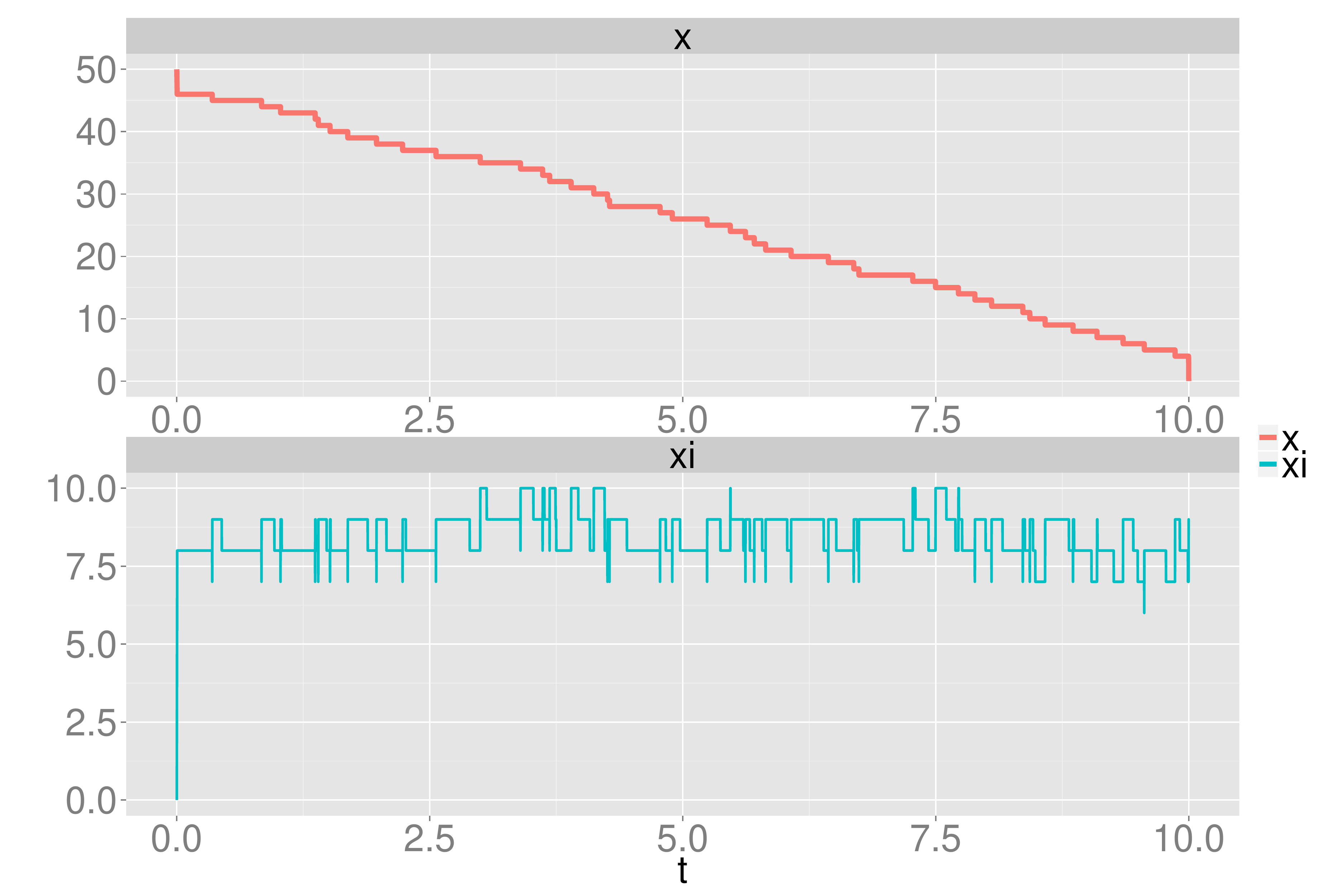}
			\caption{
			A sample path under weak recovery intensity without limit orders when $T=10$. 
			The upper and lower panels display the time evolution of the inventory $X_t$ and the degree of the market impact $\Xi_t$, respectively.
			}
			\label{fig:weak_sample_T10}
	\end{minipage}
	\hfill
	\begin{minipage}[t]{0.48\columnwidth}
		\centering
			\includegraphics[width=7.5cm]{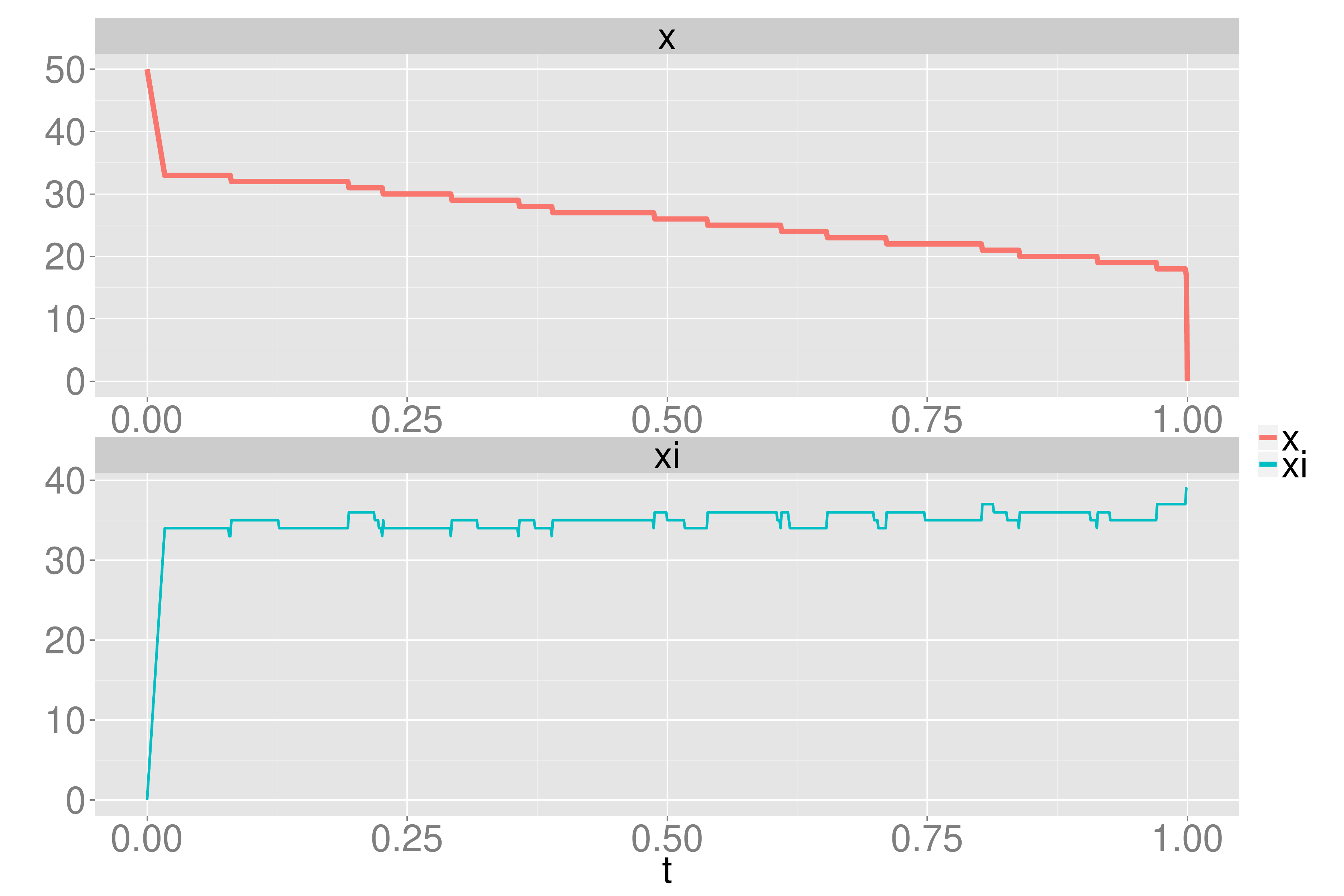}
			\caption{
			A sample path under weak recovery intensity without limit orders when $T=1$. 
			The upper and lower panels display the time evolution of the inventory $X_t$ and the degree of the market impact $\Xi_t$, respectively.
			}
			\label{fig:weak_sample_T1}
	\end{minipage}
\end{figure}

We also find that the time evolution of the tolerance follows the same pattern as in the previous subsection.
Figure \ref{fig:weak_sample_T10} shows a sample path for the current setting 
 in the same manner as in Figure \ref{fig:strong_sample_T10_fr}.
We observe that the execution of a sell order occurs when $\Xi_t$ takes a certain value, as we found previously.
Decreased tolerance is also found in the sample path,
as the value of $\Xi_t$ that triggers the execution in Figure \ref{fig:weak_sample_T10} is higher than the value in Figure \ref{fig:strong_sample_T10_fr}.
 
The most remarkable finding in this subsection is observed in Figure \ref{fig:weak_sample_T10},
where we see large trades close to the initial and terminal times,
unlike the previous case.
The large trades are described with high resolution in Figure \ref{fig:weak_sample_T1}, which shows
 a sample path with the same parameters as in Figure \ref{fig:weak_sample_T10}
 except that the terminal time is $T=1$.
These two large trades, however, each have a different structure.

Let us discuss the large trade near the initial time.
We observe that it is constructed by a sequential market order with minimum trading volume.
The sequential trades continue until $\Xi_t$ hits the trigger value for execution in the short term.
This is the result of a tradeoff between the following two desires:
(i) reducing the inventory to a proper level, which we can describe using tolerance, as quick as possible; and
(ii) avoiding an increase in the market impact $\Xi_t$ as much as possible.

The large trade at the terminal time has different characteristics.
Focusing on Figure \ref{fig:weak_sample_T1}, we see that
 (i) the slope of $X_t$ at the terminal time is steeper than the slope near the initial time; and
 (ii) $\Xi_t$ does not increase around the terminal time.
These observations imply that the large trade at the terminal time is due to the terminal condition in equation (\ref{eq:terminal_cond}).
Because we are not concerned with what happens after the terminal time, this manner of terminal execution appears to be valid.

We summarize the features of our optimal strategy obtained from the results thus far.
The strategy is made up of three components:
 (i) a large initial trade executed by sequential market orders with the minimum trade volume;
 (ii) small trades during the time period triggered by tolerance for the market impact; and
 (iii) a large terminal large trade due to the terminal condition.
The large trades appearing around the initial and terminal times vanish 
 if the execution period is long enough or if the price recovery is fast enough (see Section \ref{sec:num_res_strong}).
The tolerance for market impact plays a central role in the interpretation of the optimal behavior.

Finally, we compare our optimal strategy with those investigated in previous research that models the price recovery as a deterministic process,
 such as Obizhaeva and Wang \cite{Obizhaeva2013} .
Both strategies consist of a large initial trade, small trades during the time period and a large terminal trade.
The significant difference between them is found in the small trades.
As mentioned above, the trigger for execution, which depends on the time remaining, the inventory and the market impact,
 is at the heart of our strategy.
Because we model price recovery using stochastic processes as in equation (\ref{eq:def_xi}),
 our optimal strategy behaves dynamically.
The optimal strategies from previous studies are static in contrast.
They describe price recovery using a deterministic function where we know the exact price recovery ex ante
 and are then able to construct the execution strategy for the scheduled orders.
Our model fits this view if we substitute $\Xi_t$ into the expectation $\mathbb{E}[\Xi_t]$.
Hence, the averaged version of our strategy corresponds to the previous strategy.

\subsection{Enabling limit orders}
The waiting times found in our strategy investigated in the previous subsections lead us to include limit orders in our strategy.
Quoting the limit order appears to be efficient compared with simply waiting for the price to recover.
The parts of our model related to the limit order are indeed motivated by this view.
Recall that we include the limit order with small trade volume and a conservative evaluation of the execution price.
The approach using the limit order with small trade volume to avoid the market impact
 is similar to the iceberg strategy used by market participants in the real world.
\footnote{
For example, the NYSE offers the block reserve order which corresponds the iceberg strategy.
Academic research about the iceberg strategy is found, for instance, in Esser and M\"{o}nch \cite{Esser2007}.
} 
Hence we call our limit order strategy the quasi-iceberg strategy.
In this subsection, we discuss how the limit order is included in the optimal strategy.
Throughout this subsection we fix the parameters as follows: $\lambda^L=0.1$, $l_{\max}=3$ and $T=30$.

\begin{figure}[H]
	\centering
	\includegraphics[width=15cm]{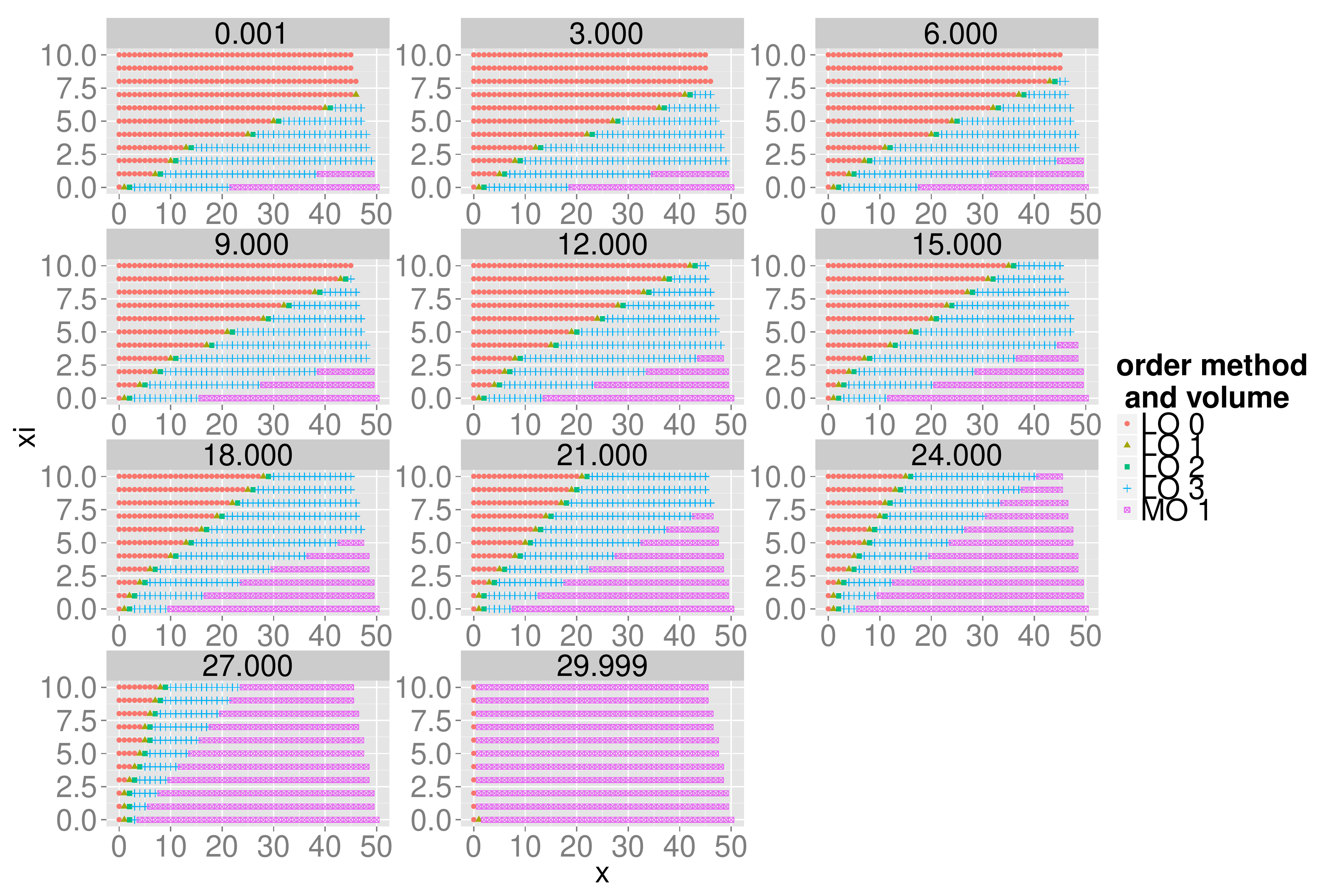}
	\caption{
		Optimal strategy with limit orders and intensity $\lambda^L=0.1$ under weak recovery intensity when $T=30$.
		The region shown is for $\Xi_t \leq 10$.
		Each panel displays a snapshot at the time indicated above the panel.
		A red dot indicates that waiting for the price to recover is optimal.
		A yellow triangle, a green square and a blue plus symbol indicate quoting a limit order with one, two or three shares is optimal, respectively.
		A purple square indicates that selling one share, the minimum traded volume, by a market order is optimal.
	}
	\label{fig:limit_stra}
\end{figure}

Figure \ref{fig:limit_stra} describes the optimal strategy in a manner similar to Figure \ref{fig:strong_t10_stra}
 although using different symbols and colors.
We find that the dynamics of the market order part of the optimal strategy follows the same rule as it does without limit orders.
We observe the new region occupied by the blue plus symbol which indicates that
 quoting a limit order with three shares is optimal.
This behavior coincides with our expectations mentioned at the beginning of the present subsection.
However, we still find that the region filled with red dots represents an optimal strategy of waiting for the price to recover.
The execution price for the limit order is evaluated conservative and hence waiting is an optimal strategy
 when frequent price recovery is expected.

We note that the regions filled with yellow triangles and green squares only appear on the boundary between
 the regions occupied by red dots, which correspond to waiting for the price to recover, and by purple squares, which correspond to a quote of $l_{\max}$.
Because there is no adverse effect due to the size of the limit order in our model,
 there is little incentive to choose trading volumes other than $l_{\max}$.
From the technical perspective, the existence of these regions is interesting because it implies that the optimal strategy changes smoothly
 from waiting to quoting a limit order with $l_{\max}$ shares.

The tolerance for market impact continues to give a powerful interpretation in the current case.
The optimal strategy has the following three phases.
In the first phase, the tolerance is high and waiting is chosen as the optimal behavior.
The tolerance then decreases and the optimal strategy changes to quoting a limit order with $l_{\max}$ shares.
In the third phase, the tolerance hits the trigger discussed in the previous subsections,
 so immediately selling one share, the minimum trading volume, by a market order is optimal.

\begin{figure}[H]
	\centering
	\includegraphics[width=15cm]{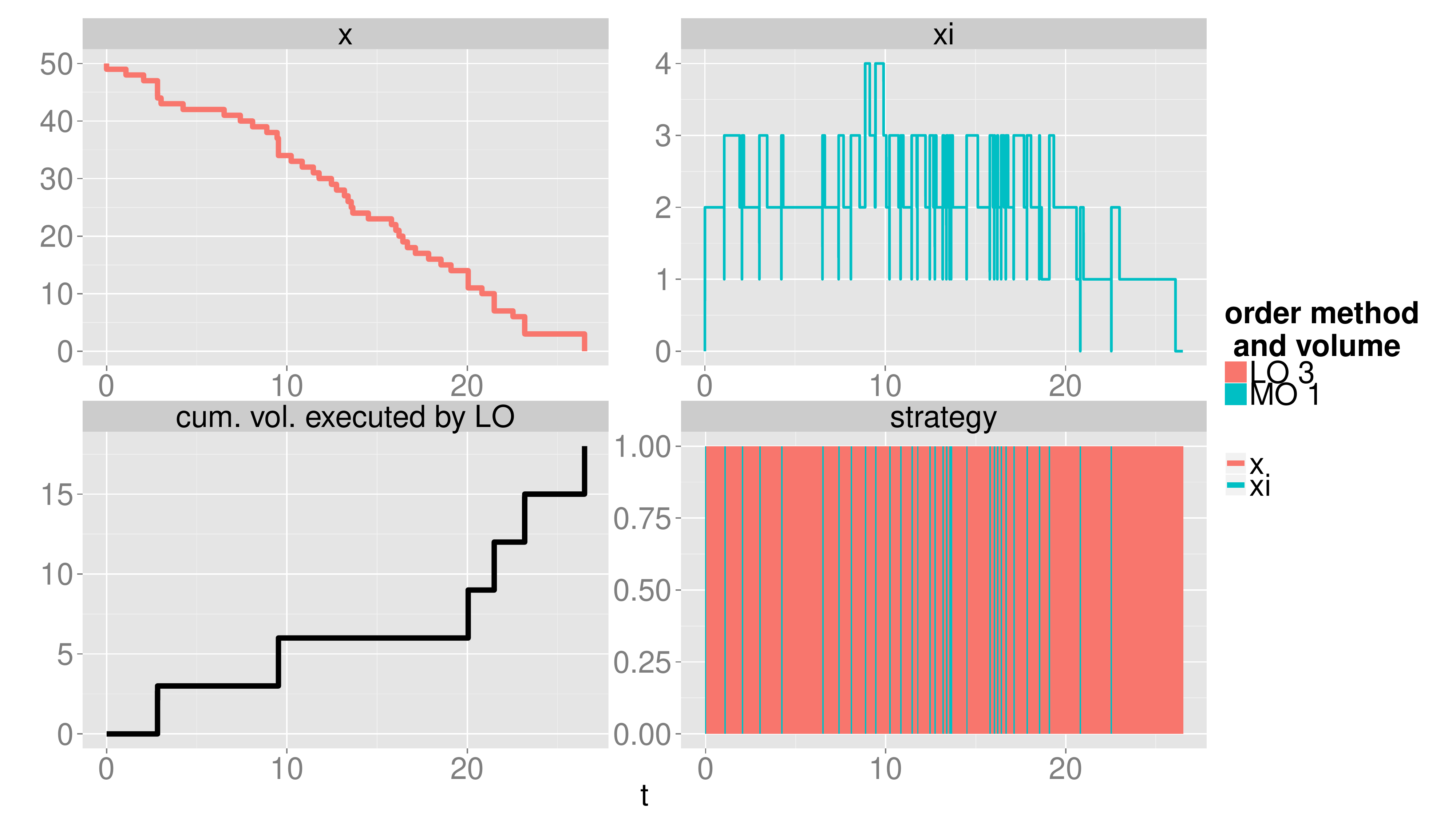}
	\caption{
			A sample path with limit orders and intensity $\lambda^L=0.1$
				under the weak recovery intensity when $T=30$. 
			The upper left and right panels display the time evolution of the inventory $X_t$ and the degree of the market impact $\Xi_t$ respectively.
			The lower left panel indicates the cumulative order volume executed by the limit orders.
			The lower right panel describes the optimal strategy:
				a blue bar indicates that immediately selling one share by a market order is optimal;
				a red bar indicates that quoting a limit order with three shares is optimal.
	}
	\label{fig:limit_sample}
\end{figure}

The strategy of waiting, however, does not appear in the investment simulation.
Figure \ref{fig:limit_sample} displays a sample path under the current settings.
In the lower right panel of the figure, which displays the optimal strategy over the time grid,
 we only observe immediately selling of one share by a market order and quoting a limit order with three shares.
This observation implies that the market order part of the optimal strategy does not allow $\Xi_t$ to exceed the trigger value, which changes the strategy to waiting.
Figure \ref{fig:limit_sample} also suggests that the dynamics of the market order part of the optimal strategy is unchanged from the discussion above.

In this sample path, the quasi-iceberg strategy works well.
The cumulative order volume executed by the limit orders is described in the lower panel of Figure \ref{fig:limit_sample}.
We observe that the limit order executions occur six times and that 18 shares are liquidated.
This liquidation due to the limit orders leads to low tolerance for the market impact found near the terminal time 
 in the upper right panel in Figure \ref{fig:limit_sample}, which indicates the time evolution of $\Xi_t$.
Another contribution of the quasi-iceberg strategy is demonstrated by the performance analysis discussed in the next subsection.

\subsection{Performance analysis}

In this subsection, we present a performance analysis of our optimal strategy.
The results in the previous subsections suggest that we are able to reduce the market impact by increasing the terminal time $T$ and
 taking a longer time span to liquidate the target asset.
On the other hand, a larger value for $T$ leads to large fluctuations from the initial price, as mentioned in the introduction.
We quantify this tradeoff through a performance analysis using the liquidation rate $R_T$ defined as follows:
\[
R_T =  \frac{ Y_T - (P_T-\Xi_T - g(X_T))X_T }{x_0 p_0}.
\]
The numerator of the above equation is the terminal wealth including the terminal execution 
 and the denominator is the wealth obtained under perfect liquidity.
Therefore, the liquidation rate is defined as the ratio between the total wealth with and without market impact.
We run the simulation under weak recovery intensity
 and evaluate the expectation and the standard deviation of the liquidation rate, $\mathbb{E}[R_T]$ and $\mathrm{SD}(R_T)$,
 for each of the terminal times given in Table \ref{tb:param_ns}.
We use geometric Brownian motion as the price process, that is, $\sigma(t,p) = \sigma p$,
 and produce $10^5$ sample paths using the parameters in Table \ref{tb:param_ns}
   as well as the parameters in Table \ref{tb:param_os}.

\begin{table}[H]
	\centering
	\begin{tabular}{|c|l|c|}
		\hline
		Parameter  & Description           & Value \\ \hline
		$T$ & terminal time & 1, 3, 5, 10, 20, 30, 50    \\
		$\sigma$ & volatility & 0.08    \\
		$p_0$ & initial price & 150 \\
		$\delta_p$ & grid size for $P_t$& 1 \\
		 $-$ & number of sample paths  & $10^5$    \\
		\hline
	\end{tabular}
	\caption{Parameters related to numerical simulation}
	\label{tb:param_ns}
\end{table}

\begin{figure}[H]
	\begin{minipage}[t]{0.48\columnwidth}
		\centering
			\includegraphics[width=7.5cm]{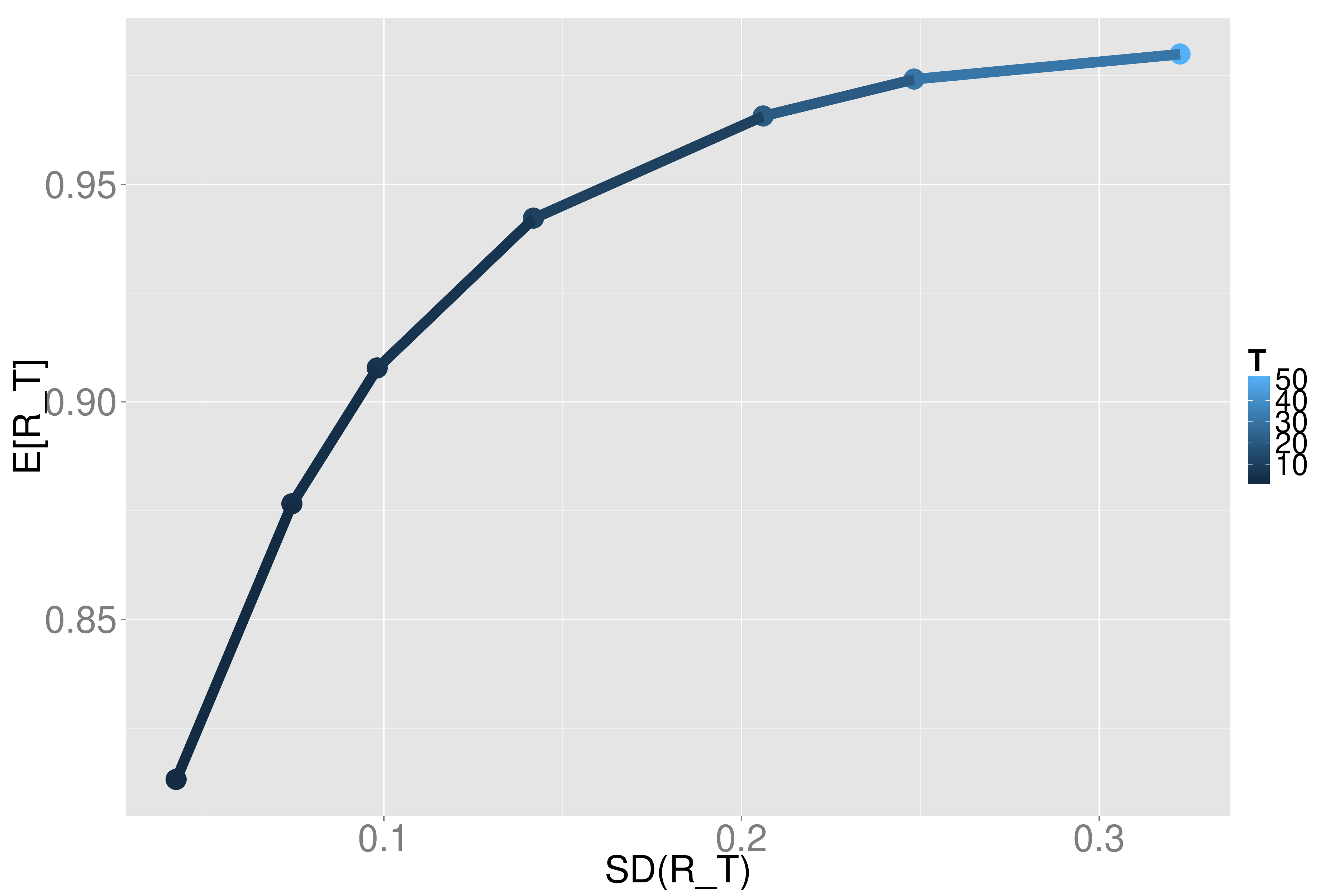}
			\caption{
				Performance graph when limit orders are prohibited.
				The expectation $\mathbb{E}[R_T]$ and the standard deviation $\mathrm{SD}(R_T)$ are plotted on 
					the vertical and horizontal axes respectively.
				The results are plotted using points colored with a lighter blue as the terminal time $T$ increases,
				and the points are interpolated linearly.
			}
			\label{fig:pa_mo}
	\end{minipage}
	\hfill
	\begin{minipage}[t]{0.48\columnwidth}
		\centering
			\includegraphics[width=7.5cm]{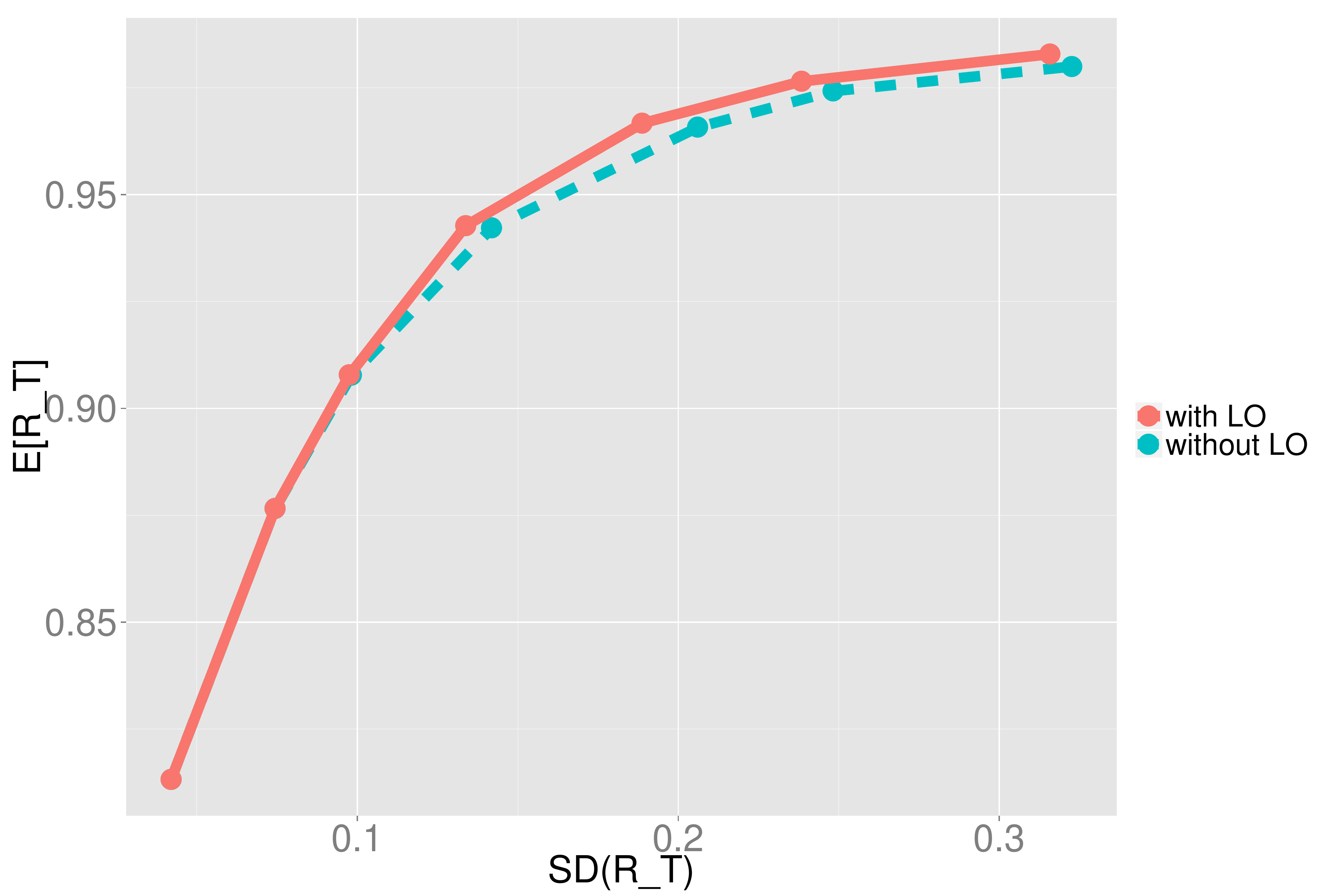}
			\caption{
				Performance graph when limit orders are permitted.
				The expectation $\mathbb{E}[R_T]$ and the standard deviation $\mathrm{SD}(R_T)$ are plotted on 
					the vertical and horizontal axes respectively.
				The dashed blue line and the solid red line describe the performance with and without the limit order, respectively.
				The blue line is the same as that displayed in Figure \ref{fig:pa_mo}.
			}
			\label{fig:pa_lo}
	\end{minipage}
\end{figure}

Let us first discuss the case that excludes the limit order.
Figure \ref{fig:pa_mo} plots $\mathbb{E}[R_T]$ and $\mathrm{SD}(R_T)$ on the vertical and horizontal axes respectively.
The points are colored with a lighter blue as $T$ increases and are interpolated linearly.
We observe that there is a monotonic increase for both $\mathbb{E}[R_T]$ and $\mathrm{SD}(R_T)$.
The former and one of the features of our strategy imply that the improvement in performance relies on reducing the market impact.
The latter is obviously caused by price fluctuation because of the large terminal time.
Hence, using $\mathbb{E}[R_T]$ and $\mathrm{SD}(R_T)$ to quantify the tradeoff
 between avoidance of market impact and price fluctuation is somewhat primitive.
The implication is obtained from the shape of the curve in Figure \ref{fig:pa_mo}.
Because the curve is concave,
 the amount of change in $\mathrm{SD}(R_T)$ required to improve $\mathbb{E}[R_T]$ a certain amount becomes larger 
as $\mathbb{E}[R_T]$ increases.
This implies that, over a long time span, liquidating the asset is not profitable.

We next consider the case where limit orders are permitted.
Recall that the limit order in our optimal strategy replaces the strategy of waiting for the price to recover
 found in the optimal strategy without the limit order.
Thus we expect an improvement in performance due to the limit order.
Figure \ref{fig:pa_lo} plots $\mathbb{E}[R_T]$ and $\mathrm{SD}(R_T)$ in the same manner as in Figure \ref{fig:pa_mo} and supports this expectation.
The dashed blue line and the solid red line in Figure \ref{fig:pa_lo} represent the results with and without the limit order respectively.
We note that the former line is the exactly same as in Figure \ref{fig:pa_mo}.
We observe that the solid red line appears on the upper left side of the dashed blue line.
This indicates that enabling limit orders leads to an improvement both in the risk and the return.
We find that the improvement stands out when $T \geq 10$.
The limit order is executed when the counterpart order arrives, 
 and the number of limit order executions declines if the terminal time is a small value.
Hence, the improvement due to the limit order is relatively small unless the terminal time is large enough.
We also find that main contribution to the improvement comes from the decrease in $\mathrm{SD}(R_T)$.
We impose a conservative evaluation on the execution value of the limit order,
 which appears to restrict the improvement in $\mathbb{E}[R_T]$.
Therefore we expect further improvements in $\mathbb{E}[R_T]$ under a more sophisticated evaluation of the limit order execution.

\section{Concluding remarks}

In this paper, we study the optimal execution problem under stochastic price recovery based on LOB dynamics.
We model the price recovery after execution of a large order by accelerating the arrival of the refilling order.
The arrival of the refilling order is defined as a Cox process whose intensity is increased by the degree of the market impact.
We include not only the market order but also the limit order in our execution strategy albeit in a restricted fashion.
Our restricted limit order strategy is regarded as a quasi-iceberg strategy.
We formulate our execution problem as a combined stochastic control problem over a finite time horizon.
The corresponding HJBQVI is solved numerically using a scheme similar to that proposed by Ieda \cite{Ieda2013}.
The performance criterion is used to maximize the terminal wealth including the terminal execution.

The optimal execution strategy without the limit order is made up of three components:
 (i) a large initial trade executed by sequential market orders with the minimum trading volume;
 (ii) unscheduled small trades during the time period; and
 (iii) a large terminal trade due to the terminal condition.
The heart of our strategy is the tolerance for the market impact, which depends on the time remaining and inventory:
  if the tolerance is high, then the optimal strategy is to wait for the recovery of the market impact,
  while if the tolerance is low, then it is optimal to liquidate by market orders.
The timing of the small trades, the second component of the strategy, is triggered by the tolerance.
Since the tolerance reflects the state at the time, our optimal strategy is dynamic.
The tolerance also governs the size of the large trades appearing around the initial and terminal times.
These trades vanish if the execution period is long enough or if the price recovery is fast enough,
 which leads to a high tolerance situation.

The intervals between the small trades appearing in the above strategy motivate including limit orders in the strategy.
Enabling limit orders under the restriction that a limit order does not affect the price dynamics,
 the strategy of waiting for the price to recover observed when limit orders are excluded is replaced by quoting the limit order.
Although we impose a conservative evaluation on the limit order execution,
 our performance analysis based on the liquidation rate, the ratio between the total wealth with and without the market impact,
 demonstrates that the improvement is caused by the inclusion of the limit order.


\bibliography{/Users/mieda/Dropbox/bibtex/library,/Users/mieda/Dropbox/bibtex/text}

\end{document}